\documentclass{aa}

\usepackage{times,float}
\usepackage{epsfig}
\usepackage{graphics}
\usepackage{amssymb}
\usepackage{float}

\begin{document}

\title{The potential of INTEGRAL for the detection of high redshift GRBs}

\titlerunning{High redshift GRBs with INTEGRAL}

{\small
 \author{J.   Gorosabel      \inst{1,2,3}
 \and    N.   Lund           \inst{2}
 \and    S.   Brandt         \inst{2}
 \and    N.J. Westergaard    \inst{2}
 \and    J.M. Castro Cer\'on \inst{3}
}

\institute{Instituto de Astrof\'{\i}sica de Andaluc\'{\i}a (CSIC), Camino Bajo de Hu\'etor, 24, E--18.008 Granada Spain.
\and       Danish Space Research Institute, Juliane Maries Vej, 30, DK--2.100 K\o benhavn \O\ Denmark.
\and       Space Telescope Science Institute, 3.700  San Martin Drive, Baltimore MD 21.218-2.463 USA.
}

\offprints{\hbox{J. Gorosabel, e-mail: {\tt jgu@iaa.es}}}

\date{Received / Accepted }

\abstract{  We discuss INTEGRAL's  capabilities to  detect a  high redshift
  population of Gamma-Ray Bursts  (GRBs) in comparison to other high-energy
  missions.  Emphasis is done on  the study of the relative capabilities of
  IBIS on board  INTEGRAL with respect to SWIFT and  HETE~2 for detecting a
  high redshift population  of GRBs.  We conclude that, if  the GRB rate is
  proportional  to the  star  formation rate,  INTEGRAL's capabilities  for
  studying GRBs  are complementary to the  ones of missions  like SWIFT and
  HETE~2, specially devoted to prompt localisations of GRBs.  Whereas SWIFT
  and HETE~2 would detect a higher number of GRBs than INTEGRAL, IBIS might
  be able  to detect  high redshift ($z  \gtrsim 7$) GRBs,  unreachable for
  SWIFT and  HETE~2.  We discuss the relevance  of performing near-infrared
  (NIR) observations of the INTEGRAL GRBs and the strategy that large class
  telescopes might follow.\keywords{gamma rays: bursts}}

\maketitle

\section{Introduction}

Gamma-Ray  Bursts (GRBs)  appear  as  brief ($10^{-3}$  s  $<$ duration  $<
1\,000$~s)  flashes of  cosmic high-energy  photons, emitting  the  bulk of
their energy above $\approx 0.1$  MeV. Among the thousands of GRBs detected
in $\gamma$-rays  since 1967 (Klebesadel  et al. \cite{Kleb73})  only $\sim
50$  GRBs have  been  identified at  optical wavelengths\footnote{See  {\tt
http://www.mpe.mpg.de/$\sim$jcg/grbgen.html}.}.     For   $36$    of   them
spectroscopic redshifts have been measured, ranging from $z=0.0085$ (Galama
et al.  \cite{Gala98}) to $z=4.50$ (Andersen et al.  \cite{Ande00}).

  A canonical model has emerged  for powering long GRBs (durations $\gtrsim
2$~s): collapse  of a  massive star onto  a black  hole.  This leads  to an
intense  flash of $\gamma$-ray  photons followed  by an  expanding fireball
which emits  radiation at  lower frequencies (Woosley  \cite{Woos93}).  The
spectroscopic  association of GRB~030329  with the  type I$_{c}$  SN 2003dh
supports strongly  this framework (Hjorth et al.   \cite{Hjor03}; Stanek et
al. \cite{Stan03}).   Short GRBs (durations $\lesssim 2$~s)  have not shown
yet   a  conclusive   optical   counterpart  (see   Castro-Tirado  et   al.
\cite{Cast02} for a further discussion).

We know  today that  most long duration  GRBs originate  at $z >  0.5$.  In
fact,  it is  thought that  their intrinsic  brightness would  allow  us to
detect  these  events at  epochs  corresponding  to  the formation  of  the
earliest stellar  populations.  Thus, they may  be used as  probes into the
first  stages of  star formation  and their  spectra may  reveal  the early
heavy-element enrichment of the  interstellar medium (ISM). GRBs originated
by  the  exploding  primitive  population  of stars  are  suggested  to  be
detectable up to redshifts $z \sim 30$ (Lamb \& Reichart \cite{Lamb00}).

ESA's INTEGRAL  satellite offers unique  capabilities for the  detection of
GRBs  thanks   to  its  high   sensitivity  and  imaging   capabilities  at
$\gamma$-rays,  $X$-rays and  optical  frequencies. INTEGRAL  is the  first
$\gamma$-ray spacecraft that combines imaging instruments of high precision
and a continuous  real time telemetry link.  Developments  in the GRB field
over the  past few  years have made  increasingly clear that  INTEGRAL, and
specially  the  IBIS\footnote{Imager  on  board  the  INTEGRAL  satellite.}
$\gamma$-ray  imager, could make  a significant  contribution. IBIS  is the
coded mask  telescope on  board INTEGRAL dedicated  to imaging over  a wide
($15$ keV  $- 10$ MeV) energy  range (Ubertini et  al.  \cite{Uber03}).  To
date, neither of the  three classical INTEGRAL GRBs (considering GRB~031203
as an  $X$-ray flash) with  optical counterparts identified have  had their
redshift  measured  (G\"otz et  al.   \cite{Gotz03};  Castro-Tirado et  al.
\cite{Cast03}; Masetti et al. \cite{Mase04}).

In  Sect.~\ref{integral}  we  perform  a  rough  comparison  of  INTEGRAL's
sensitivity    with   respect   to    other   $\gamma$-ray    and   $X$-ray
missions-instruments.   Sect.~\ref{faintgrbs} studies  in  more detail  the
capabilities  of   the  most  sensitive   high-energy  missions-instruments
(INTEGRAL/IBIS,  HETE~2/WXM and  SWIFT/BAT) to  detect high  redshift GRBs.
Further,  in  Sect.~\ref{IBIS-SWIFT}  we  focus on  the  INTEGRAL/IBIS  and
SWIFT/BAT  relative number  of high  redshift GRB  detections.  As  an ado,
Sect.~\ref{nir} makes emphasis on  the relevance of the near-infrared (NIR)
observations    to     detect    high    redshift     afterglows.     Last,
Sect.~\ref{conclusions} presents the conclusions of our study.

\section{Sensitivity comparisons: INTEGRAL expectations}
\label{integral}

In  Table~\ref{table1} we  provide the  main  performances of  a number  of
different  space missions/instruments with  capabilities for  GRB research.
The  missions/instruments are  divided  in two  groups  depending on  their
energy  range.   In  the  first  group  the  INTEGRAL/IBIS  sensitivity  is
normalized  with respect to  the Burst  Alert Telescope  (BAT) $\gamma$-ray
instrument on board SWIFT.\footnote{Launch scheduled for the second half of
2004. Hereafter, when  we refer to SWIFT  we will focus only  on BAT (which
provides  the GRB  triggers for  the mission)  and not  on the  $X$-ray and
$UV$/Optical Telescopes on board SWIFT. See Gehrels et al.  (\cite{Gehr04})
for     further     details.}       In     the     second     group     the
INTEGRAL/JEM-X\footnote{Joint  European  Monitor  for  $X$-rays,  on  board
INTEGRAL.}   and SAX/WFC\footnote{Wide  Field Cameras  on  board BeppoSAX.}
rough sensitivities are given  compared to the HETE~2/WXM\footnote{The Wide
$X$-ray Monitor is the most sensitive instrument on board HETE~2.  The rest
of the  HETE~2 instrumentation will  not be considered.}  one.   Because we
concentrate on  determining the relative  number of detections  for several
missions/instruments (see  Sect.  \ref{IBIS-SWIFT})  we will not  perform a
dedicated  calculation of the  individual instrumental  sensitivities; that
goes beyond the  scope of our study (see Band  \cite{Band03} for a detailed
analysis).

\begin{table*}[t]
\caption{\label{table1} Theoretical estimations of several missions-instruments capabilities.}
\begin{center}
\begin{tabular}{lccccc}

\hline
Mission/Instrument &  Area    &  Coverage          & Energy band &   Orbit &Relative   \\
                   &  cm$^2$  &  $\%$ of $4\pi$ str. &  (keV) &   efficiency &sensitivity$^{\dagger}$\\

\hline
\hline

SWIFT/BAT           & 5200 & $14^{\star}$     &  15 to 150& 0.6 & 1.0\\

INTEGRAL/IBIS (ISGRI)& 3000 & $0.9^{\star}$    &  15 to 150& 0.8 & 3.0\\   

\hline
\hline

 HETE~2/WXM           & 360 &   13 & 2 to 25& 0.5 & 1.0\\
                                                
 INTEGRAL/JEM-X & 1000&  0.1 & 2 to 35& 0.8 & 19.0\\   

 SAX/WFC        & 530 ($\times$2)& 2 ($\times$2) & 2 to 30 &  0.5 & 3.1\\

\hline
\hline
\multicolumn{5}{l}{$\star$ Half coded field of view (FOV).}\\
\multicolumn{5}{l}{$\dagger$ Estimate based on the $P_{ins} \sim \sqrt{\Omega/A}$
recipe.}\\
\hline
\end{tabular}
\end{center}
\end{table*}

  In  order  to  compare  the  sensitivities  of  different  missions  Band
  (\cite{Band03}) made emphasis on the  need of expressing them in terms of
  a common  energy band.  Band  (\cite{Band03}) determined the  peak photon
  flux  threshold  (hereafter named  $P_{ins}$,  measured  in ph  cm$^{-2}$
  s$^{-1}$),  for  several  detectors,  including HETE~2/WXM,  SAX-WFC  and
  SWIFT/BAT.  INTEGRAL/IBIS and INTEGRAL/JEM-X  were not considered in that
  study.   In  Band  (\cite{Band03})   all  the  detection  thresholds  are
  normalised to the $1-1\,000$  keV energy band, making several assumptions
  about  the detectors'  properties (accumulation  time,  detector response
  matrix,   background   model)   and   GRBs'   spectra.\footnote{In   Band
  (\cite{Band03}) the GRB spectra are  described by the Band function (Band
  et al. \cite{Band93}), where the photon number flux is given by:

\[
 N(E)=
  \left\{ \begin{array}{ll}
   N_0 (\frac {E}{100 keV})^{-\alpha_1}\exp(-\frac{E}{E_0})& 
                                       \mbox{$E \leq E_b$}\\
   N_0 (\frac {E_b}{100 keV})^{\alpha_2-\alpha_1} (\frac {E}{100
  keV})^{-\alpha_2} \exp(\alpha_1-\alpha_2)  & 
                                       \mbox{$E > E_b$ }
  \end{array} \right.
\]

where $E_b=(\alpha_2  - \alpha_1)E_0$, and  $E_0$ determines the  maximum of
$N(E)E^2$  given by $E_p=(2-\alpha_1)E_0$.   $\alpha_1$ and  $\alpha_2$ are
the photon spectral indexes at frequencies below and above $E_b$.}

Based  on  Figs.~5~and~7  in  Band  (\cite{Band03})  we  assumed  detection
thresholds of $P^{(1-1\,000) {\rm  keV}}_{\rm HETE~2/WXM} = 4$ ph cm$^{-2}$
s$^{-1}$ for HETE~2/WXM and  $P^{(1-1\,000) {\rm keV}}_{\rm SWIFT/BAT} = 2$
ph cm$^{-2}$ s$^{-1}$ for SWIFT/BAT.   These two thresholds are valid for a
broad range of  $E_p$ values of around several hundreds  of keV, (a typical
range for GRBs) and refer to  the $1-1\,000$ keV energy band.  We note that
$E_p$  is  the  spectral  maximum  in  the  $\nu  F_{\nu}$  (or  equivalent
$N(E)E^{2}$)  representation.  Many  studies have  reported results  in the
$50-300$ keV band, mainly because this band corresponded to the sensitivity
range  of  BATSE\footnote{The Burst  And  Transient  Source Experiment;  it
operated  on  board the  Compton  Gamma-Ray  Observatory  between 1991  and
2000.}.  Thus, in the present work  we decided to carry out the sensitivity
comparisons in a common $50-300$ keV reference energy band. The election of
another energy band is arbitrary and it would not affect our results, since
it would only  introduce a constant multiplicative factor  on the $P_{ins}$
values  calculated   for  each  mission-instrument  (as  well   as  on  the
luminosity, $L$, as it will be shown in equation 2).}

Setting  $\alpha_1  = \alpha_2  =  1.5$  (see  Sect.  \ref{faintgrbs}),  we
calculated a  flux ratio $P^{(50-300) {\rm  keV}}/P^{(1-1\,000) {\rm keV}}=
0.09$.   For  a  power  law  spectrum this  detection  threshold  ratio  is
independent of $E_p$ (which would correspond to a constant straight line in
Fig.~1 of Band \cite{Band03}).   By transforming the assumed $1-1\,000$ keV
thresholds,  the  derived $50-300$  keV  sensitivities  for HETE~2/WXM  and
SWIFT/BAT  become  $P^{(50-300)  {\rm  keV}}_{\rm HETE~2/WXM}  =  0.35$  ph
cm$^{-2}$ s$^{-1}$  and $P^{(50-300) {\rm keV}}_{\rm SWIFT/BAT}  = 0.17$ ph
cm$^{-2}$ s$^{-1}$,  respectively.  We  note that those  sensitivity values
refer to the peak flux.

Estimates carried  out for  the IBAS\footnote{INTEGRAL Burst  Alert System}
sensitivity yield a threshold of $\sim 0.14-0.22$ ph cm$^{-2}$ s$^{-1}$ for
INTEGRAL/IBIS (the ISGRI part) in  the $20-200$ keV energy band (Mereghetti
et al.   \cite{Mere03}).  Hereafter we assume  a conservative INTEGRAL/IBIS
sensitivity limit  of $P^{(20-200) {\rm  keV}}_{\rm INTEGRAL/IBIS}=0.22$ ph
cm$^{-2}$  s$^{-1}$,  which  corresponds  to $P^{(50-300)  {\rm  keV}}_{\rm
INTEGRAL/IBIS}  =  0.12$  ph   cm$^{-2}$  s$^{-1}$,  assuming  $\alpha_1  =
\alpha_2= 1.5$.   Therefore, the INTEGRAL/IBIS capabilities  to detect high
redshift GRBs relative to  SWIFT/BAT (presented in Sect.  \ref{IBIS-SWIFT})
have to be considered as a pessimistic estimate of the actual INTEGRAL/IBIS
potential.

An alternative,  simplified way to  verify the photon peak  flux thresholds
for  the   instruments  not   considered  by  Band   (\cite{Band03})  (like
INTEGRAL/IBIS and  INTEGRAL/JEM-X) would be to assume  that the sensitivity
threshold $P_{ins}$  for a  GRB is proportional  to the square-root  of the
background count rate and inversely  proportional to the square-root of the
detector  area.   Thus we  have  $P_{ins}  \propto \sqrt{\Omega/A}$,  where
$\Omega$ is the sky coverage and $A$ is the detector area.  This assumption
would  yield  a  sensitivity  ratio  of  $3.0$  between  INTEGRAL/IBIS  and
SWIFT/BAT   (see  Table~\ref{table1})  and,   therefore,  a   threshold  of
$P^{(50-300) {\rm keV}}_{\rm INTEGRAL/IBIS} = 0.06$ ph cm$^{-2}$ s$^{-1}$.

In  principle,   the  most  reliable   comparison  that  can  be   done  in
Table~\ref{table1} is between SWIFT/BAT and INTEGRAL/IBIS (the ISGRI part),
because they  are based  on very similar  detector technologies  (CdZnTe in
SWIFT/BAT, CdTe  in INTEGRAL/IBIS) and  share a similar energy  band.  This
simple  estimate  is a  factor  of two  lower  than  the $P^{(50-300)  {\rm
keV}}_{\rm INTEGRAL/IBIS}  = 0.12$ ph cm$^{-2}$  s$^{-1}$ threshold assumed
for  INTEGRAL/IBIS.  Hence,  it supports  the use  of this  threshold  as a
conservative upper limit of the real INTEGRAL/IBIS sensitivity.

From the above estimates it is evident that INTEGRAL/IBIS will be the
most sensitive  GRB detector (at least in  the 15 $-$ 150  keV energy band)
ever flown  and not likely  to be matched,  sensitivity wise, by  any other
mission within the coming decade.

Rescaling   the   HETE~2/WXM    threshold   with   the   $P_{ins}   \propto
\sqrt{\Omega/A}$  recipe  we  obtain  a sensitivity  of  $P^{(50-300)  {\rm
keV}}_{\rm  INTEGRAL/JEM-X}  = 0.02$  ph  cm$^{-2}$s$^{-1}$.  However,  the
difference  in  detector technologies  of  INTEGRAL/JEM-X  with respect  to
HETE~2/WXM  makes   this  number  an  uncertain  estimate   of  the  actual
INTEGRAL/JEM-X  sensitivity.   Furthermore,  the on  flight  INTEGRAL/JEM-X
performances  have been  changed during  the first  months of  the INTEGRAL
mission, so its real sensitivity is well above $0.02$ ph cm$^{-2}$s$^{-1}$.
Besides, the  reduced number of  bursts that INTEGRAL/JEM-X will  detect (a
very few a  year) does not support performing  a specific calculation aimed
at studying its capabilities for high redshift bursts.

On  the   other  hand  an  estimate   based  on  Figs.~3  and   5  of  Band
(\cite{Band03})  yields a  relative sensitivity  of  $\sim 3  - 4$  between
HETE~2-WXM  and SAX-WFC.   This  is  in agreement  with  the HETE~2/WXM  vs
SAX/WFC  relative  sensitivity  estimate  given  by  the  $P_{ins}  \propto
\sqrt{\Omega/A}$ expression (see Table~\ref{table1}).

\section{ Detectability of a high redshift population of GRBs}
\label{faintgrbs}

We  have   selected  the  most   sensitive  missions-instruments  (present:
  INTEGRAL/IBIS,  HETE~2/WXM,  and future:  SWIFT/BAT)  to calculate  their
  capabilities  of detecting  a  high redshift  population  of bursts.   To
  estimate the  number of GRBs that these  missions-instruments will detect
  we assume that:

\begin{itemize}  
  
\item  GRB   spectra  can   be  described  by   power  laws   ($F_\nu  \sim
  \nu^{-\alpha}$).  After exploring  the impact  of $\alpha$  on  the final
  results (which do  not change qualitatively) and for  simplicity, we have
  assumed  a  value of  $\alpha=1.5$,  typical  of  GRBs (van  Paradijs  et
  al. \cite{Vanp00}).
  
\item The GRB rate is proportional  to the star formation rate (SFR) in the
  Universe.   The  SFR  considered  is  the  one  given  by  Rowan-Robinson
  (\cite{Rowa99},  \cite{Rowa01}) for  $z <  5$ and  the one  calculated by
  Gnedin \& Ostriker (\cite{Gned97}) for $z \ge 5$ (see Fig.~\ref{Fig1}).

\begin{figure}[t]
 \centering 
   \resizebox{\hsize}{!} {\includegraphics{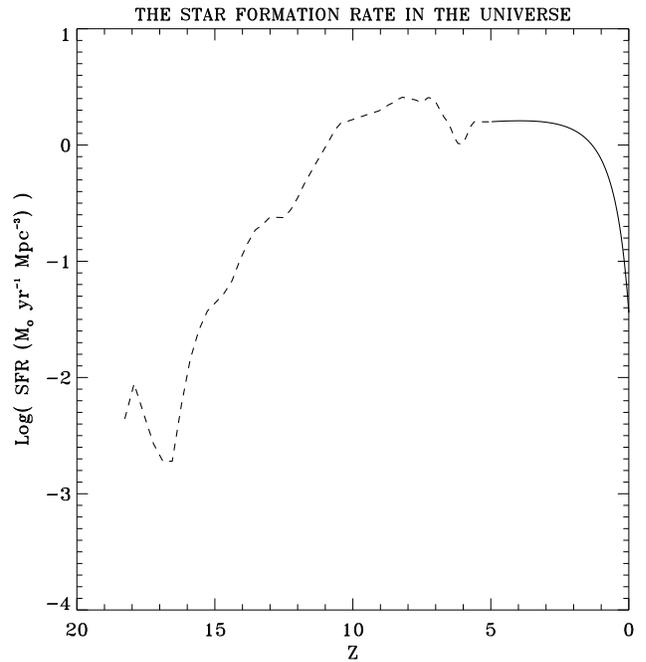}}
  \caption{\label{Fig1} The plot shows  the SFR in the Universe as a 
    function of the  redshift.  The dashed line represents  the SFR derived
    from  numerical   simulations  for  $z  \ge  5$   (Gnedin  \&  Ostriker
    \cite{Gned97}).  The  solid line shows the  SFR at the $z  < 5$ region
    based   on  observational   estimates   (Rowan-Robinson  \cite{Rowa99},
    \cite{Rowa01}).   The  transition  between  the two  regions  has  been
    smoothed.}
\end{figure}

\item The GRB peak (isotropic) photon luminosity function is given by:

\[ 
 S(L)=
  \left\{ \begin{array}{ll}
           L^{\beta}
      & \mbox{  $L_{min} < L < L_{max}$} \\
    0 & \mbox{  Otherwise } 
  \end{array} \right.
\]

being $L$  the peak photon  luminosity and $\beta$ the  luminosity function
index.   $L_{min}$,  $L_{max}$  determine   the  width  of  the  luminosity
function. We have assumed a value of $\beta=-1$.

\item Although the  effect of several Universe models  has been checked, we
  choose to  use the  most popular cosmological  parameters in  this paper:
  $\Omega_m=0.3$,   $\Omega_{\Lambda}=0.7$,  H$_{o}   =  65$   km  s$^{-1}$
  Mpc$^{-1}$.

\end{itemize}

\begin{figure}[t]
\centering 
   \resizebox{\hsize}{!} {\includegraphics{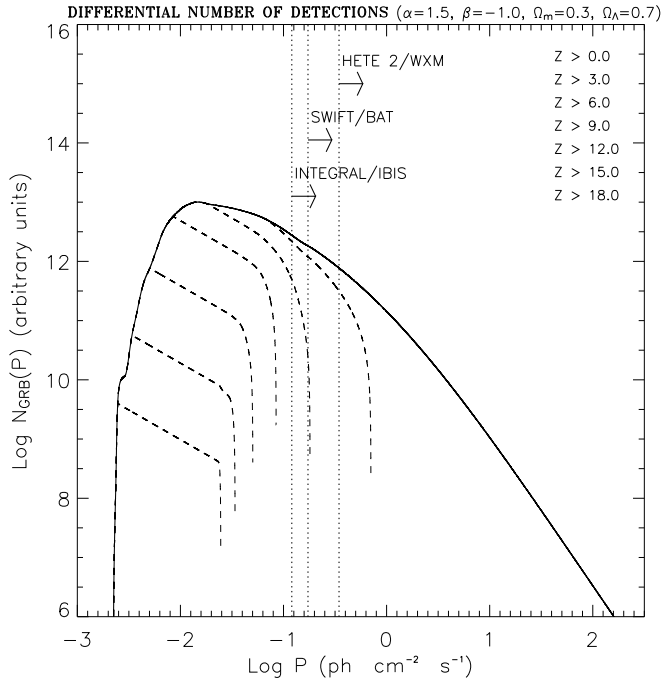}}

\caption{\label{Fig2} Differential  peak photon flux  distribution of GRBs.
  The solid curve  shows the differential peak photon  flux distribution if
  all  redshifts are  considered,  i.e., $N_{GRB}(P)$.   The dashed  curves
  represent  the differential peak  photon flux  distribution of  GRBs when
  only  GRBs   with  $z   >  z_{edge}$  are   taken  into   account,  i.e.,
  $N_{GRB}(P,z_{edge})$. The  intention of this  figure is to show  how the
  Gaussian  like  solid  curve  displaces  leftwards (low  $P$  values)  as
  $z_{edge}$ increases.  The  dashed lines plotted in the  figure are shown
  for the purpose of the example, using arbitrary values of $z_{edge}$, not
  marked  for simplicity.   As $z_{edge}$  increases the  closer  GRBs (and
  therefore larger  $P$ bursts)  are left out  (right tail of  the Gaussian
  like   distribution).   The  vertical   lines  represent   the  detection
  thresholds for the different missions-instruments, showing the arrows the
  detectability region.}
\end{figure}

For the  above assumptions the differential  GRB detection rate  at a given
peak photon flux  $P$ in a detector (ph cm$^{-2}$s$^{-1}$)  is given by the
following convolution integral:

\begin{equation}
$$~~~~~~~~~~~~~~~~~~~~N_{GRB}(P) = C ~\Omega~\epsilon  \int_{0}^{\infty} R_{GRB} S(L) dL $$~,
\end{equation}

\noindent where $\epsilon$ is the  efficiency of the orbit, $\Omega$ is the
instrumental coverage of the sky and $R_{GRB}$ is the GRB detection rate if
they were standard candles,  i.e., $R_{GRB}= \frac{\rm SFR({\it z})}{(1+z)}
\frac{dV(z)}{dz} \frac{dz}{dP}$, being $V$  the comoving volume.  The value
of the  proportionality constant $C$ is  unknown.  Fig.~\ref{Fig2} displays
$N_{GRB}(P)$  as well as  the detection  thresholds of  several high-energy
missions-instruments.

 The  relationship  between $L$,  $z$ and  $P$   is given by  the following
expression:

\begin{equation}
 ~~~~~~~~~~~~~~~~~~~~~~~~~~~P= \frac{L}{4 \pi D(z)^2 (1+z)^{\alpha} }~,
\end{equation}

\noindent  where $D(z)$  is  the comoving  distance.   In our  calculations
different    values   of    $\alpha$,   L$_{min}$,    L$_{max}$,   $\beta$,
$\Omega_{\Lambda}$ and $\Omega_{m}$ are considered. Variations of $\alpha$,
$\beta$, $\Omega_{\Lambda}$ and $\Omega_{m}$ do not change the final result
qualitatively.  On  the other hand, the variations  of $L_{min}$, $L_{max}$
are  more  relevant  for  determining  the  number  of  high  redshift  GRB
detections.  We consider  the pessimistic  case of  a relatively  faint and
narrow  luminosity  function   $S(L)$  defined  by  $L_{min}=10^{57.5}$  ph
s$^{-1}$ and $L_{max}=10^{58.5}$ ph  s$^{-1}$.  The $L_{min}$ and $L_{max}$
values used in our study correspond  to the narrowest $S(L)$ among the ones
used by Lamb \& Reichart  (\cite{Lamb00}), and consistent with the observed
peak photon luminosity distribution (the  measured $S(L)$ is at least $1.7$
times  wider;  Stern et  al.   \cite{Ster02})\footnote{  Our $L_{min}$  and
$L_{max}$  values are  also in  agreement  with the  empirical peak  photon
luminosity function used as  reference by Lamb \& Reichart (\cite{Lamb00});
$\log L =  58.1 \pm 0.7$}.  The assumption of a  wider $S(L)$ would broaden
the   Gaussian  like   curves  (both   solid  and   dashed)   displayed  in
Fig.~\ref{Fig2}, extending  their tails to higher $P$  values and therefore
enhancing the number of GRBs detected at very high redshift.

We can calculate the contribution to  (1) by GRBs with redshift larger than
$z_{edge}$ (see dashed curves of Fig.~\ref{Fig2}), using:

$$N_{GRB}(P,z_{edge}) = C~\Omega~\epsilon \int_{0}^{\infty} H(z(L),z_{edge}) R_{GRB} S(L)dL~,$$

\noindent where $H(z(L),z_{edge})$ is  a step function that vanishes unless
$z(L)~>~z_{edge}$.     Obviously,   $N_{GRB}(P)   =    N_{GRB}(P,0)$,   and
$\frac{N_{GRB}(P,z_{edge})  }{ N_{GRB}(P)}~\le~1$.  Last, we  can calculate
the number  of GRBs  detected above a  given instrumental peak  photon flux
threshold $P_{ins}$ that have redshifts larger than $z_{edge}$:

$$N_{GRB}(P_{ins},z_{edge})=\int_{P_{ins}}^{\infty} N_{GRB}(P,z_{edge})dP~.$$

\begin{figure}[t]
\centering
   \resizebox{\hsize}{!} {\includegraphics{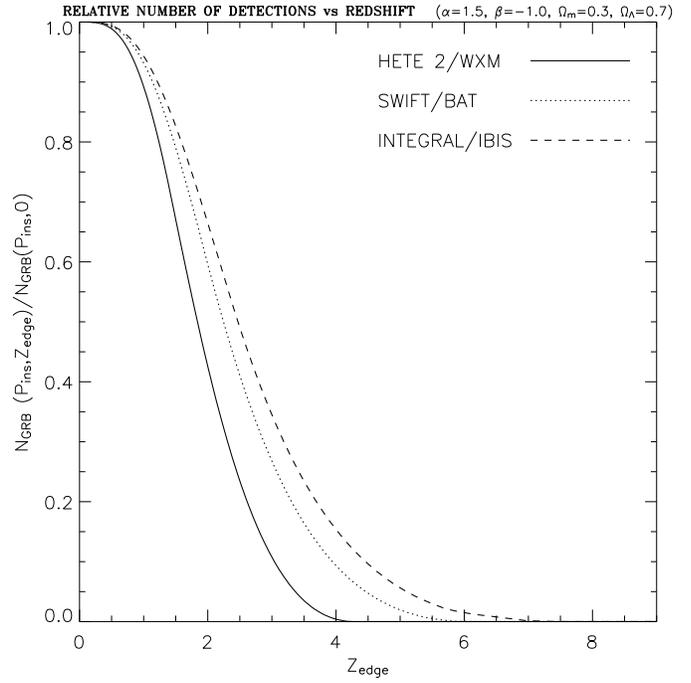}}
\caption{\label{Fig3} Relative  number of detections  as a function  of the
  redshift.  This plot shows, for several high-energy missions-instruments,
  the  fraction of  the  detected GRBs  that  have a  redshift larger  than
  $z_{edge}$.}
\end{figure}

Not knowing the  proportionality constant $C$ of (1), we  can not derive an
absolute value for  $N_{GRB}(P_{ins},z_{edge})$.  However, we can determine
the        relative        quantity        $\frac{N_{GRB}(P_{ins},z_{edge})
}{N_{GRB}(P_{ins},0)}$, which provides us the proportion of detections that
have a redshift larger than $z_{edge}$ (see Fig.~\ref{Fig3}).

\section{INTEGRAL/IBIS {\scriptsize vs} SWIFT/BAT; comparison  of the 
number of GRB detections}
\label{IBIS-SWIFT}

Fig.~\ref{Fig3}  shows   the  relative  number  of   detections  (given  by
$\frac{N_{GRB}(P_{ins},z_{edge})}{N_{GRB}(P_{ins},0)}$),  as a  function of
the redshift, for HETE~2/WXM, SWIFT/BAT and INTEGRAL/IBIS.  As it is shown,
$\sim15\%$  of the  GRBs detected  by  INTEGRAL/IBIS will  have a  redshift
larger than $4$.  For SWIFT/BAT the  $z > 4$ population will be $\sim 10\%$
of the total  number of detections.  In the case  of INTEGRAL/IBIS the tail
of $\frac{N_{GRB}(P_{ins},z_{edge})}{  N_{GRB}(P_{ins},0)}$ extends even up
to redshifts  of $z_{edge}  > 7$.  SWIFT/BAT  and HETE~2/WXM will  detect a
closer population of bursts,  specially HETE~2/WXM.  HETE~2/WXM is the less
sensitive GRB mission, being constrained to detect bursts with redshifts $z
< 4.3$.  This prediction is in agreement with the maximum redshift measured
for a HETE~2/WXM GRB.\footnote{Among  the 12 HETE~2/WXM GRBs with confirmed
spectroscopic  redshifts  to date,  the  maximum  redshift  is reached  for
GRB~030323 at  $z=3.372$ (Vreeswijk et al.   \cite{Vree04}).}  Therefore we
will not consider  HETE~2/WXM in the following study,  aimed at calculating
the relative  number of GRB detections  as a function of  the redshift.  We
will  concentrate on  comparing INTEGRAL/IBIS  and  SWIFT/BAT capabilities.
Besides,  as we  previously noted,  the similar  energy range  and detector
technologies of INTEGRAL/IBIS and SWIFT/BAT suggest a reliable comparison.

For determining the  relative number of detections between  two missions, A
and B, the following calculation has to be performed:

\begin{equation}
$$~f_{\rm A-B}(z_{edge})={\frac{\Omega_{\rm A}
    \epsilon_{\rm A}\int_{P_{\rm A}}^{\infty}\int_{0}^{\infty}
    H(z(L),z_{edge})R_{GRB}S(L)dLdP}{\Omega_{\rm B} \epsilon_{\rm B} \int_{P_{\rm B}}^{\infty}\int_{0}^{\infty}H(z(L),z_{edge})R_{GRB}S(L)dLdP}}$$~.
\end{equation}

This function  will give the  relative number of  GRB detections with  $z >
z_{edge}$.   We have  applied (2)  to  derive the  detection ratio  between
INTEGRAL/IBIS and  SWIFT/BAT as a  function of the GRB  population redshift
($f_{{\rm INTEGRAL/IBIS}-{\rm SWIFT/BAT}}(z_{edge})$) .

\begin{figure}[t]
  \centering \resizebox{\hsize}{!}  {\includegraphics{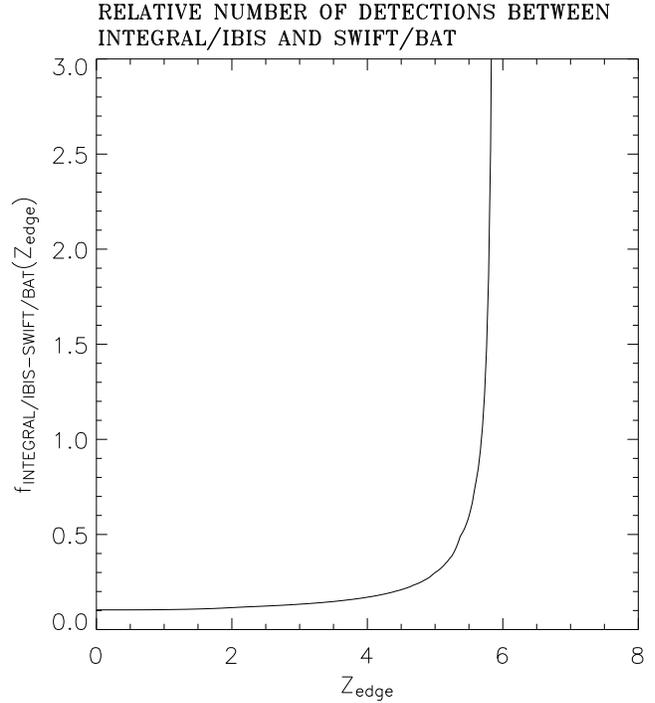}}
 \caption{\label{Fig4}  The   plot  shows   the  GRB  detection   ratio  of
  INTEGRAL/IBIS  with  respect  to  SWIFT/BAT  as a  function  of  the  GRB
  population   redshift.    $f_{   {\rm  INTEGRAL/IBIS}-{\rm   SWIFT/BAT}}$
  increases  with  redshift because  of  the  better  sensitivity for  high
  redshift GRBs of INTEGRAL/IBIS.  At redshifts $z \lesssim 5.6$ the larger
  field of  view of SWIFT  dominates, making $f_{  {\rm INTEGRAL/IBIS}-{\rm
  SWIFT/BAT}}< 1$.}
\end{figure}

As  it  is  shown  in   Fig.~\ref{Fig4},  for  $z_{edge}  <  5.6$  $f_{{\rm
INTEGRAL/IBIS}-{\rm SWIFT/BAT}} < 1$.  In other words, at low redshifts the
large field  of view  (FOV) of SWIFT/BAT,  in comparison  to INTEGRAL/IBIS,
dominates the  number of  detections.  On the  other hand, for  $z_{edge} >
5.6$, INTEGRAL/IBIS sensitivity becomes  the dominating factor and $f_{{\rm
INTEGRAL/IBIS}-{\rm SWIFT/BAT}} > 1$.  Thus,  from the point of view of the
ground based  strategy of detecting $z  \gtrsim 6$ afterglows,  it might be
more  efficient to  observe INTEGRAL/IBIS  GRBs than  to  observe SWIFT/BAT
bursts.  If we are interested in using GRBs to study the reionisation epoch
that occurred at $z\sim7$ (Loeb \& Barkana \cite{Loeb01}), then it would be
relevant to prioritise the follow up of INTEGRAL/IBIS GRBs.  Nonetheless we
emphasise  that INTEGRAL/IBIS high  redshift detectability  predictions are
based  on   low  number  statistics,   so  they  are  subjected   to  large
fluctuations.

The  self consistency  of  our procedure  can  be checked  by studying  the
prediction of expression (3) for the particular case when $z_{edge}=0$.  If
we consider $z_{edge}=0$, then $f_{{\rm INTEGRAL/IBIS}-{\rm SWIFT/BAT}}(0)$
gives us  the fraction of GRBs  detected with $z>0$,  i.e., considering all
the detections  independently of  their redshifts.  We  predict a  value of
$f_{{\rm INTEGRAL/IBIS}-{\rm  SWIFT/BAT}}(0) =  1/9.7$ for the  fraction of
the total  number of GRBs  detected by INTEGRAL/IBIS compared  to SWIFT/BAT
(see  Fig.~\ref{Fig4}  at  $z_{edge}=0$).   The number  of  GRBs  currently
detected  by INTEGRAL/IBIS is  $\sim 1$  GRB per  month (Mereghetti  et al.
\cite{Mere03}),  whereas  the  last  updated estimates  for  the  SWIFT/BAT
detection rate yield  $\sim 110$ GRBs per year  (Heyl \cite{Heyl03}).  This
gives  a fraction  of $\sim  1/9.2$  for the  number of  detected GRBs,  in
agreement with our prediction.

Although INTEGRAL/JEM-X FOV and sensitivity are less suitable than the ones
of INTEGRAL/IBIS  to detect  GRBs, the spectral  peak of the  high redshift
GRBs  (usually  at  $100-1200$ keV)  will  be  in  the detection  range  of
INTEGRAL/JEM-X.    So  given   that  INTEGRAL/JEM-X   is   co-aligned  with
INTEGRAL/IBIS,  it might  be also  useful to  detect the  redshifted prompt
$\gamma$-ray component.

\section{NIR observations for INTEGRAL GRBs}
\label{nir}

The most significant inconvenience to  follow up INTEGRAL GRB afterglows is
the  large fraction of  the mission's  time devoted  to scan  the extincted
Galactic  centre/plane.  An  additional problem  (not specific  to INTEGRAL
like  the previous one)  for detecting  high redshift  GRBs comes  from the
Ly-$\alpha$  blanketing  effect  that  strongly  attenuates  the  radiation
observed  at $\lambda  <  1216  \times (1+z)~$\AA.  Both  drawbacks can  be
mitigated if ground based searches are carried out in the NIR, specially in
the $K$-band, where $i)$ the Galactic extinction drops by a factor $\sim 6$
compared to  the optical, and  $ii)$ the Ly-$\alpha$ blanketing  problem is
eliminated up to very high redshifts ($z\sim17$).

There are further  advantages to searching for high  redshift afterglows in
the NIR. Because of their fading behaviour (typically, GRB afterglows decay
following a  power law of the  form $F_\nu \sim t^{-\delta}$,  where $t$ is
the time since  the onset of the $\gamma$-ray event  and $\delta$ the decay
index) the time dilation effect increases the observed flux at a fixed time
of observation  after the  GRB.  Thus current  and future ground  based NIR
facilities  could be  able  to  detect high  redshift  afterglows that  are
invisible in the optical bands.

\begin{figure}[t]
\begin{center}
  {\includegraphics[width=\hsize]{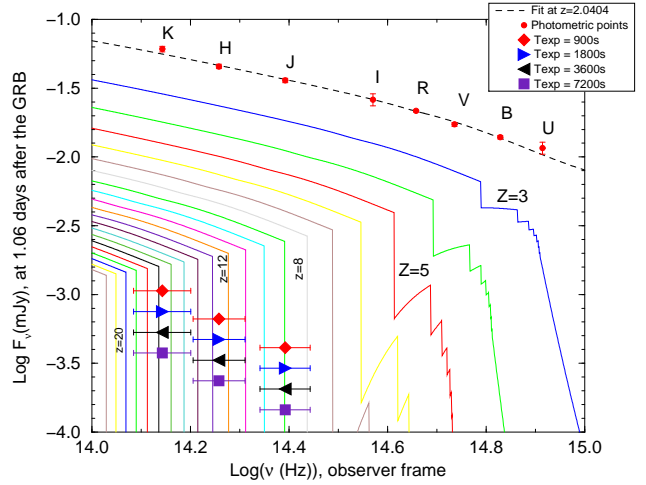}}
  \caption{\label{301c} The  plot shows the evolution of  the observed flux
    of a  typical GRB  afterglow (GRB~000301c) when  it is  redshifted from
    $z=2.0404$ (spectroscopic redshift; Jensen et al.  \cite{Jens01}) up to
    $z=22$.  The circles show the photometric measurements of the afterglow
    from the  $U$ to the  $K$-band, taken $1.06$  days after the  GRB.  The
    dashed line  represents the SED  fitted to the photometric  points (see
    main text).   The rhomboids, triangles and squares  show the $JHK$-band
    sensitivities (5$\sigma$) foreseen for  the future GTC 10m telescope (+
    EMIR  instrument), for  several  exposure  times.  As  it  can be  seen
    afterglows can be  detected in the three $JHK$ bands  up to $z=9$, with
    only a exposure time of  900~s per filter.  $K$-band counterparts would
    be detectable up to $z\sim17$.  For $T_{exp} >$ 900~s the detectability
    of the GRB~000301c  afterglow with the GTC (+EMIR)  is dominated by the
    Ly-$\alpha$  blanketing effect,  and  is basically  independent of  the
    exposure time employed in the observations.  The Ly-$\alpha$ blanketing
    absorption has been modeled following Madau (\cite{Mada95}).}
\end{center}
\end{figure}

As an example, in Fig.~\ref{301c} we display the very well sampled spectral
energy  distribution (SED)  of  the GRB~000301c  afterglow  (Jensen et  al.
\cite{Jens01})  for different  redshifts, once  the  Ly-$\alpha$ blanketing
absorption has  been modeled  (Madau \cite{Mada95}). The  photometric point
flux  densities ($F_\nu$;  see dashed  line in  Fig.~\ref{301c})  have been
fitted following  an expression of  the form $F_\nu \sim  \nu{^\eta} \times
10^{-0.4 A_\nu}$,  where $\eta$ is  the afterglow power law  spectral index
and $A_\nu$  is the  absorption in the  GRB host  galaxy at a  frequency of
$\nu$.   $A_\nu$ has  been  parameterised  in terms  of  $A_V$ following  a
typical SMC  extinction law (Pei \cite{Pei92}).   For illustration purposes
we have overplotted, for different exposure times, the $JHK$-band 5$\sigma$
sensitivities  foreseen  for  the  Espectr\'ografo  Multiobjeto  Infrarrojo
(EMIR) (Balcells  \cite{Balc98}), currently being  build for the 10  m Gran
Telescopio    Canarias    (GTC)    (Rodr\'{\i}guez    Espinosa    et    al.
\cite{Rodr98})\footnote{For  additional   information,  please,  visit  the
following sites: {\tt http://www.ucm.es/info/emir/index\_e.html}, and also:
{\tt http://www.gtc.iac.es/}.}.

We would like  to emphasise several points. First,  this simple redshifting
exercise  does not  require  any  assumption on  the  star formation  rate,
luminosity  function, etc\dots  Second, the  GRB~000301c afterglow  was not
specially bright,  so an intense afterglow  (like GRB~990123; Castro-Tirado
et al.  \cite{Cast99}) could be detectable even at higher redshifts. Third,
the photometric points displayed in  the figure correspond to a fairly late
epoch observation (they were acquired $\sim1.06$ days after the burst, time
measured in the observer's frame), so afterglows observed a few hours (even
minutes) after  the burst would  be reachable by smaller  facilities (i.e.,
NIR   robotic   telescopes  like   REM;   Zerbi   et  al.    \cite{Zerb01},
\cite{Zerb03}).

As it  is shown  in Fig.~\ref{301c}, rapid  NIR observations  of afterglows
performed with  10 m  class telescopes (e.g.,  the GTC equipped  with EMIR)
would  be dominated  by the  Ly-$\alpha$ blanketing  effect since,  above a
given exposure time threshold ($T_{exp} \sim 900$~s for the particular case
of GRB~000301c and the GTC),  the detectability is basically independent of
the exposure time employed in the observations.

At  high redshifts  the observer  would be  sampling the  $UV$ part  of the
afterglow synchrotron spectrum where  the extinction of the GRB environment
is expected  to be severe.  Thus one  of the mayor problems  to predict the
NIR detectability of a high redshift  afterglow would come from the, so far
unknown,  extinction law  and dust  content describing  both the  local GRB
environment and  the large scale line  of sight properties  within the host
galaxy.  Furthermore, the $UV$ opacity  in the environment close to the GRB
progenitor  might  be  time  dependent.   Hence,  a  detailed  quantitative
prediction  of the number  of afterglows  detected in  the NIR  (beyond the
scope  of  this  work)   would  require  modeling  the  physical  processes
describing the $UV$ absorption of the afterglow radiation and its evolution
at different scales (from several AU to kpc) around the GRB progenitor.

\section{Discussion and conclusions}
\label{conclusions}

 In   the   present   paper   we   study  the   capabilities   of   several
missions-instruments to  detect high redshift  GRBs. At low  and moderately
high redshifts ($0  <z \lesssim 5.6$) the large  FOV of SWIFT/BAT, compared
to  the instrumentation  on board  INTEGRAL,  dominates the  number of  GRB
detections.   Nonetheless, the  better sensitivity  of IBIS  makes INTEGRAL
more efficient detecting GRBs beyond $z\sim5.6$.

 In the most popular cosmological  models, the first sources of light began
 at a redshift of $z\eqsim30$ and reionised most of the Universe by $z \sim
 7$ (Loeb  \& Barkana  \cite{Loeb01}).  Thus the  study of  the Ly-$\alpha$
 forest,  present in  the  high redshift  INTEGRAL/IBIS afterglow  spectra,
 might  constrain  the  epoch   of  such  reionisation  (Gunn  \&  Peterson
 \cite{Gunn65}), probing  the ionisation state of  the intergalactic medium
 (IGM) as a function of redshift.   According to our estimates this kind of
 studies  will be  less  productive if  SWIFT/BAT  (even more  HETE~2/WXM),
 rather than INTEGRAL/IBIS, GRBs are followed up.

 Further,  optical/NIR  spectroscopy  of  high redshift  afterglows  would
 provide  additional  information on  the  primitive  IGM  and ISM.   High
 resolution spectroscopy  of high  redshift afterglows might  reveal metal
 absorption lines, which  could trace the metal enrichment  history of the
 Universe.  In  principle this is the  same technique as the  one used for
 studying the damped Ly-$\alpha$ systems  (DLAs), which are located in the
 line of sight of quasars.   However GRBs are cleaner probes than quasars,
 because a  brief phenomenon such as a  GRB would not modify  the state of
 the environment at large distances  as quasars do (the Str\"omgren sphere
 of  high  redshift  quasars  is  of  the  order  of  Mpc;  White  et  al.
 \cite{Whit03}).  GRBs would allow to study the ISM of basically unaltered
 host galaxies,  highly ionised if they  had harboured a  quasar during at
 least one  million years.   Moreover, GRBs are  brighter (although  for a
 short time) than quasars, so GRBs can probe the ISM of dustier galaxies.

 Theoretical studies show that the  birthrate of Population {\rm III} stars
 produces a  peak in the  SFR in the  Universe at redshifts $16  \lesssim z
 \lesssim 20$, while the birthrate  of Population {\rm II} stars produces a
 much  larger and  broader peak  at redshifts  $2 \lesssim  z  \lesssim 10$
 (Valageas \& Silk \cite{Vala99}).  If GRBs are produced by the collapse of
 massive stars,  they are  expected to occur  at least  at $z \sim  10$ and
 possibly up to  $z \sim 15-20$.  The mere detection  of very high redshift
 GRBs would give us for the first time direct confirmation of the existence
 of the earliest stellar generations.

 Rapid NIR observations are a complementary strategy to study high redshift
 INTEGRAL/IBIS  GRBs.   $K$-band observations  performed  with current  and
 planned  ground  based  NIR  facilities   might  detect  GRBs  up  to  the
 theoretical  limit  imposed  by  the Ly-$\alpha$  blanketing  ($z\sim17$).
 Realistically however the $z\sim17$ upper  limit ought to be decreased due
 to possible $UV$ absorption present in the GRB host galaxy.

 In  conclusion, the  INTEGRAL/IBIS's  capabilities for  studying GRBs  are
 complementary  to  the ones  of  missions-instruments  like SWIFT/BAT  and
 HETE~2/WXM, specially  devoted to  prompt localizations of  GRBs.  Whereas
 SWIFT/BAT and  HETE~2/WXM would detect  more GRBs than  INTEGRAL/IBIS, the
 latter might detect high redshift GRBs unreachable to the earlier.  Future
 works  might invert  the logical  outflow followed  in the  present paper.
 Once spectroscopic redshifts have been measured for a large sample of GRBs
 (and therefore  the number of detected  GRBs as a function  of redshift is
 known),  the  equations could  be  inverted in  order  to  obtain the  SFR
 evolution with  redshift. INTEGRAL,  and specially IBIS,  could be  a very
 valuable tool to trace the SFR rate in the early Universe.

\section*{Acknowledgments}
 We thank our anonymous referee for fruitful and constructive comments.  We
are  very  grateful to  David  L.  Band  for  helpful  information used  to
determine the instrumental thresholds.


\begin{thebibliography}{}

\bibitem[2000]{Ande00} Andersen, M.I., Hjorth, J., Pedersen, H., et al. 2000, A\&A, 364, L54.

\bibitem[1998]{Balc98} Balcells, M., 1998, Ap\&SS, 263, 361.

\bibitem[1993]{Band93} Band, D.L., Matteson, J., Ford, L., et al. 1993, ApJ, 413, 281.

\bibitem[2003]{Band03}Band, D.L., 2003, ApJ, 588, 945.

\bibitem[1999]{Cast99} Castro-Tirado, A.J., Zapatero-Osorio, M.R., Caon,
  N., et al. 1999, Science, 283, 2069.

\bibitem[2002]{Cast02} Castro-Tirado, A.J., Castro Cer\'on, J.M., Gorosabel,
  J., et al. 2002, A\&A, 393, L55.

\bibitem[2003]{Cast03} Castro-Tirado, A.J., Gorosabel, J., Guziy, S., et
  al. 2003, A\&A, 411, L315.

\bibitem[1998]{Gala98} Galama, T.J., Vreeswijk, P.M., van Paradijs, J., et
  al. 1998, Nature, 395, 670.

\bibitem[2004]{Gehr04}  Gehrels,  N., Chincarini,  G., Giommi, P.,  et al.
  2004, ApJ, in press, {\tt [astro-ph/0405233]}.

\bibitem[1997]{Gned97} Gnedin, N.Y.,  \& Ostriker, J.P., 1997, ApJ, 486, 581.

\bibitem[2003]{Gotz03} G\"otz, D., Mereghetti, S., Hurley, K., et al. 2003,  A\&A,  409, 831.

\bibitem[1965]{Gunn65} Gunn, J.E., \& Peterson, B.A., 1965, ApJ, 142, 1633.

\bibitem[2003]{Heyl03} Heyl, J.S., 2003, ApJ, 592, 401.

\bibitem[2003]{Hjor03} Hjorth, J., Sollerman, J., M\o ller, P., et al. 2003, Nature, 423, 847.

\bibitem[2001]{Jens01} Jensen, B.L., Fynbo, J.P.U., Gorosabel, J., et al. 2001, A\&A, 370, 909.

\bibitem[1973]{Kleb73} Klebesadel, R.W., Strong, I.B., \&, Olson, R.A., 1973, ApJ, 182, L85.

\bibitem[2000]{Lamb00} Lamb, D.Q., \& Reichart, D.E., 2000, ApJ, 536, L1. 

\bibitem[2001]{Loeb01} Loeb, A., \& Barkana, R., 2001, ARA\&A, 39, 19.

\bibitem[1995]{Mada95} Madau, P., 1995, ApJ, 441, 18.

\bibitem[2004]{Mase04} Masetti, N., Palazzi, E., Rol, E., et al. 2004, GCN Circ. \# 2515.

\bibitem[2003]{Mere03} Mereghetti, S., G\"otz, D., Borkowski, J., Walter,
  R., \& Pedersen, H., 2003, A\&A, 411, L291.

\bibitem[1992]{Pei92} Pei, Y.C., 1992, ApJ, 395, 130.

\bibitem[1998]{Rodr98} Rodr\'{\i}guez Espinosa, J.M., \'Alvarez, P., \&
  S\'anchez, F., 1998,  Ap\&SS, 263, 355.

\bibitem[1999]{Rowa99} Rowan-Robinson, M., 1999, Ap\&SS,  266, 291.

\bibitem[2001]{Rowa01} Rowan-Robinson, M., 2001, ApJ, 549, 745.

\bibitem[2003]{Stan03} Stanek, K.Z., Matheson, T., Garnavich, P.M., et al. 2003, ApJ, 591, L17.

\bibitem[2002]{Ster02} Stern, B.E., Tikhomirova, Ya., \& Svensson, R., 2002, ApJ,  573, 75.

\bibitem[2003]{Uber03}  Ubertini, P.,  Lebrun, F.,  Di Cocco,  G.,  et al. 2003, A\&A, 411, L131.

\bibitem[1999] {Vala99} Valageas, P., \& Silk, J., 1999, A\&A, 347, 1.

\bibitem[2000]{Vanp00} van Paradijs, J., Kouveliotou, C., \& Wijers, R.A.M.J., 2000, ARA\&A, 38, 379.

\bibitem[2004]{Vree04}Vreeswijk, P.M., Ellison, S.L., Ledoux, C., et al. 2004, A\&A, 419, 927.

\bibitem[2003]{Whit03} White, R.L., Becker, R.H., Fan, X., \& Strauss, M.A., 2003, AJ, 126, 1.

\bibitem[1993]{Woos93} Woosley, S.E., 1993, ApJ, 405, 273.

\bibitem[2001]{Zerb01}Zerbi, F.M., Chincarini, G., Ghisellini, G., et al. 2001, AN,  322, 275.

\bibitem[2003]{Zerb03}Zerbi, F.M., Chincarini, G., Ghisellini, G., et al. 2003, SPIE, 4841, 737.

\end{thebibliography}
\end{document}